\documentclass[pra,twocolumn,showpacs]{revtex4}

\usepackage[SquareTraceBrackets]{quantum}
\usepackage{graphicx,bm,natbib,upgreek}

\usepackage[T1]{fontenc}   
\usepackage{fouriernc}
\usepackage[scaled]{helvet}

\newcommand{\micro}{$\upmu$}

\begin{document}

\title{\large\textsf{\textbf{Cluster state generation with ageing qubits}}}

\author{Joshua M. Waddington, Joshua A. Kettlewell and Pieter Kok}\email{p.kok@sheffield.ac.uk}
\affiliation{Department of Physics \& Astronomy, University of Sheffield, Sheffield S3 7RH, United Kingdom}

\begin{abstract}
 \noindent Cluster states for measurement-based quantum computing can be created with entangling operations of arbitrary low success probability. This will place a lower bound on the minimum lifetime of the qubits used in the implementation. Here, we present a simple scaling argument for the $T_2$ time of the qubits for various cluster state geometries, and we estimate the size of the qubit reservoir that is needed for the creation of mini-clusters.   
\end{abstract}

\date{\today}
\pacs{03.67.Bg, 03.67.Pp, 03.67.Lx, 42.50.Dv}

\maketitle

\section{Introduction}
\noindent
One of the most promising approaches to building a quantum computer is the one-way model, or measurement-based quantum computing \cite{raussendorf01,raussendorf03}. In this model, the computation proceeds by performing measurements on single qubits that are part of a highly entangled cluster state. This architecture has comparatively high fault-tolerance thresholds \cite{raussendorf06,raussendorf07,Li10,fujii10,monroe12} and is therefore attractive for the practical implementation of a quantum computer. Cluster states can be created with a variety of entangling gates \cite{browne05,barrett05,lim05,lim06,benjamin06,moehring07}. The gates  are typically probabilistic, for example when post-selection is used to increase the fidelity of the gate operation. It was shown that with arbitrary low probability entangling gates, universal quantum computation can be achieved theoretically \cite{barrett05,campbell08}, but for practical implementations multi-qubit nodes are highly desired \cite{benjamin06,campbell07,nickerson12}. Matsuzaki, Benjamin and Fitzsimons studied the accumulation of errors in the creation of cluster states when the success probability of the two-qubit gate is small \cite{matsuzaki10}. The resource-efficient creation of cluster states was studied in detail by Gross, Kieling and Eisert \cite{gross06}.

\begin{figure}[b]
 \includegraphics[width=8.6cm]{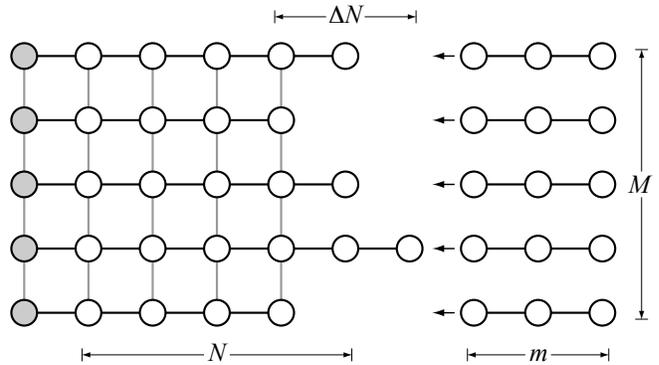}\medskip
 \caption{``Just-in-time'' creation of cluster states for meas\-ure\-ment-based quantum computing. The shaded qubits on the left are measured to drive the computation forward, while mini-clusters of size $m$ are added probabilistically on the right. The buffer size is $N$, and the variance in the buffer size is $\Delta N$. The number of logical qubits participating in the computation is $M$.}
 \label{fig:mbqc}
\end{figure}

While from a theoretical perspective it is standard to first construct the entire cluster state for the computation, this has the obvious drawback that the qubits which partake in the computation last will be ageing for the duration of the entire computation. Consequently, they will have ample time to decohere. A solution to this problem is the ``just-in-time'' creation of cluster states, shown in Fig.~\ref{fig:mbqc}. At each time step $\Delta t$ the computation proceeds by measuring the shaded qubits on the left-hand side of the cluster state, and new mini-clusters of size $m$ are added to the cluster state on the right-hand side (or, at least an attempt is made to add the mini-clusters to the cluster state). The buffer size of the cluster is denoted by $N$, and the variability of the buffer size due to the probabilistic addition of new qubits to the cluster is $\Delta N$.  The time step $\Delta t$ is long enough to accommodate one attempt at the entangling procedure or one qubit measurement (whichever takes longer to complete). The number of logical qubits partaking in the computation is $M$, and all operations that can be performed in parallel are assumed to do so.

In any practical implementation of a quantum computer, qubits will inevitably degrade over time. This is not only true for matter qubits such as ions, NV centres, and quantum dots, but also for photonic qubits that need to be stored in a quantum memory during feed-forward processing. The relevant time scale that measures the coherence time of the qubit is the $T_2$ time, and for quantum computation to work $T_2$ must be much longer than the time it takes from the initial qubit initialisation to the measurement that drives the computation. In other words, the qubits must reach the measurement stage in time before they decohere too much. In the following, we derive a lower bound on the $T_2$ time based on very general considerations about the just-in-time method for cluster state generation. We also give an estimate of the size $Q$ of the off-line reservoir of qubits that is needed to create on average one mini-cluster of size $m$, given constraints on the time the qubits spend in entangled cluster states. The size of the reservoir $Q$ scales exponentially with the inverse success probability $p$ of the entangling procedure, and a reasonably high $p$ is required for the practical implementation of cluster state quantum computing.  

\section{One-Dimensional Clusters}
\noindent
Cluster states that are universal for quantum computation are regular graphs of at least two dimensions. However, for simplicity we first consider one-dimensional cluster states. In Fig.~\ref{fig:mbqc} this can be understood by removing the grey vertical edges in the cluster state. Higher-dimensional cluster states are considered in Sec.~\ref{sec:hd}. 

There are many different entangling procedures, which behave slightly differently in creating cluster states. In general, a successful entangling gate may involve the measurement of one or more qubits from the cluster, or it may result in so-called ``cherries''---qubits that are not part of the main length of the mini-cluster. Similarly a failed entangling gate often requires us to purify the cluster by measuring a subset of qubits. If $c_1$ is the number of qubits that do not contribute to the main buffer length in the case of a successful entangling gate, and $c_2$ is the number of qubits that are removed from the buffer length after a failed entangling gate, the average growth rate $R$ per time step $\Delta t$ of the linear cluster is 
\begin{align}
 R = p(m-c_1) - (1-p) c_2-1\, ,
\end{align}
where $p$ is the success probability of the entangling procedure. We also included the subtraction of the qubit that is measured for the computation. The just-in-time method requires $R=0$, which leads to the size of the mini-cluster
\begin{align}\label{eq:minicluster}
 m = \frac{1+p c_1 +(1-p)c_2}{p}\, .
\end{align}
A list of values of $c_1$ and $c_2$ for various entangling procedures is given in Table~\ref{tab:entgate}.

Next, we need to establish how long it takes to create mini-clusters of size $m$. This (average) time $\tau$ is the age of the clusters before they are added to the main cluster, and it depends strongly on the \emph{additional} resources, such as the total reservoir of qubits and the ability to address arbitrary qubits at each time step. For example, if we have an infinite reservoir of qubits we can create a mini-cluster of size $m$ in 2 steps. The first step creates entangled qubit pairs, while the second step attempts to entangle these pairs in longer strings. since we operate all entangling gates in parallel there will be sufficient instances where all entangling procedures are successful. 

The average age of the qubit at the time of the final measurement can now be determined. We add the time it took to create the mini-cluster of length $m$ to the average buffer length $\braket{N}$. Since the last qubit in the mini-cluster takes the longest to reach the measurement stage, we also add the length $m$ of the mini-cluster:
\begin{align}
 \tau_{\rm av} = \tau + (\braket{N} + m) \Delta t\, , 
\end{align} 
This qubit age must be only a small fraction of the $T_2$ time. We can therefore write (in units of $\Delta t$)
\begin{align}\label{eq:t2}
 T_2 = \alpha ( \tau + \braket{N} + m) \, . 
\end{align} 
The factor $\alpha\gg 1$ is determined by the fault-tolerance threshold for the quantum computer architecture that is employed. We shall therefore refer to $\alpha$ as the \emph{fault-tolerance factor}, the precise value of which depends on the details of the error correction protocols.

\begin{table}[t]
\begin{ruledtabular}
\begin{tabular}{p{4cm}p{1cm}p{1cm}}
 Entangling Procedure 	& $c_1$ & $c_2$ \smallskip\cr \hline
 Type-I fusion 			& 2 & 2 \cr
 Type-II fusion 			& 3 & 2 \cr
 Double-heralding 		& 1 & 1 \cr
 Repeat-until-success 	& 1 & 0 \cr
 Broker-client model 		& 0 & 0 \cr
\end{tabular}
\caption{Characteristics of various entangling procedures for cluster state generation. The constants $c_1$ and $c_2$ indicate how many qubits shorter the main buffer length becomes after a successful and a failed entangling attempt, respectively.}
\label{tab:entgate}
\end{ruledtabular}
\end{table}

Since the growth of the buffer is a probabilistic process, for increasingly lengthy calculations there will be a growing chance that the buffer is entirely consumed by the qubit measurements. This is a catastrophic error, the probability of which must be below the fault tolerance threshold. This is accomplished by creating a buffer that is large enough, and leads to another dimensionless parameter, called the \emph{buffer factor} $\beta$. It multiplies the fluctuations in the buffer length:
\begin{align}\label{eq:buffer}
 \braket{N} = \beta \Delta N \, .
\end{align}
The size of the fluctuations can be determined by a classical random walk argument, and is given by
\begin{align}\label{eq:randomwalk}
 (\Delta N)^2 = \frac{(1+c_2)^2}{p} - c_2(2-c_2)\, .
\end{align}
The buffer factor $\beta$ will depend on the duration of the computation, as well as the number of logical qubits. It is proportional to the logarithm of these quantities.

The buffer factor $\beta$ can be related to the fault tolerance factor $\alpha$. When the buffer runs out, the logical qubit associated with that string  will experience a heralded dephasing error. The probability of such an error must fall below the fault tolerance threshold, which we set at $\alpha^{-1} = \tau_{\rm av}/T_2$. Assuming an approximately Gaussian  random walk, we find that
\begin{align}
 \beta = \sqrt{2\ln\alpha}\, .
\end{align}

\begin{figure}[t!]
 \includegraphics[width=8.5cm]{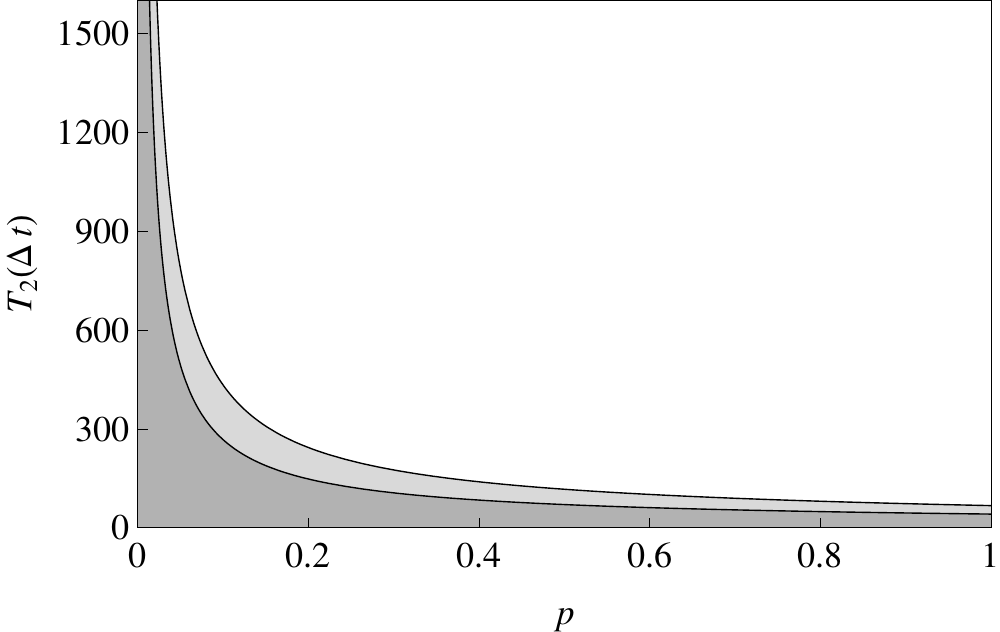}
 \caption{Threshold of the $T_2$ lifetime of the qubits in the linear cluster state as a function of the entangling gate success probability $p$ [c.f.\  Eq.~(\ref{eq:threshold})] with $\alpha=10$ for double heralding (upper curve) and the broker-client model (lower curve) \cite{benjamin06}. The shaded region is where qubits age too fast for sustainable quantum computation. There is no lower limit (i.e., a threshold) to the success probability $p$.}
 \label{fig:t2}
\end{figure}

We can now combine Eqs.~(\ref{eq:minicluster}), (\ref{eq:t2}),  (\ref{eq:buffer}),  and (\ref{eq:randomwalk}) to obtain the minimum $T_2$ time for the qubits as a function of the quantum computer architecture characteristics:
\begin{align}\label{eq:threshold}
 T_2 & = \alpha (\tau + \beta \Delta N + m) \\
 & = \alpha \left( \tau + \beta\sqrt{\frac{(1+c_2)^2}{p} - c_2(2-c_2)} + \frac{1+p c_1 +(1-p)c_2}{p} \right) \cr
 & = \alpha\left( \tau + \sqrt{\frac{(8-2p)\ln\alpha}{p}} + \frac{2}{p} \right) \quad\text{for}~ c_1=c_2=1\, .\nonumber
\end{align}
Since we do not want the production time of the mini-clusters to be the bottle neck in the $T_2$ time, we require that $\tau$ scales as the highest power of $1/p$ in Eq.~(\ref{eq:threshold}). For $\alpha=10$, and a conservative estimate of $\tau = 1/p$, the threshold for $T_2$ is given in Fig.~\ref{fig:t2}. The shaded region under the curve indicates that the $T_2$ time is too short for creating sufficiently pure cluster states with an entangling gate of success probability $p$.

\section{Higher-Dimensional Clusters}\label{sec:hd}
\noindent
As mentioned in the previous section, one-dimensional cluster states are not sufficient for efficient universal quantum computation. We will have to use the linear clusters, which each represent a single logical qubit in the computation, and create edges between them. However, we must not jeopardise the continuous chain of physical qubits that constitute the logical qubit, since we took great pains to create the chain in the first place. We can shorten the chain, but we must not break it.

\begin{figure}[b]
 \includegraphics[width=8.5cm]{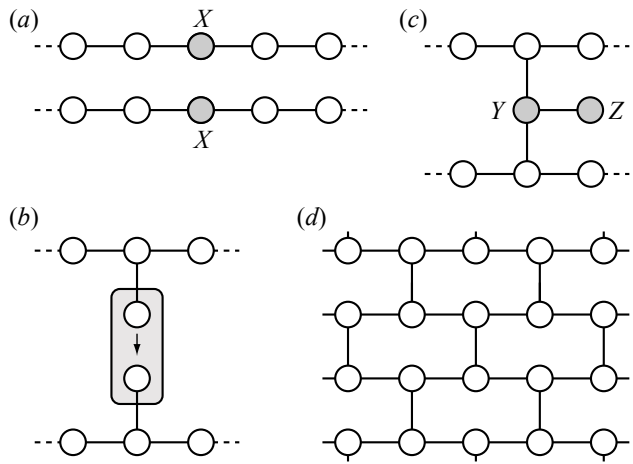}
 \caption{Vertical edges in the cluster state can be created without jeopardising the horizontal edges as follows: ($a$) Perform an $X$ measurement on two qubits; ($b$) attempt the entangling gate on the resulting cherries. Upon failure, remove the cherries with a $Z$ measurement and return to the previous step; ($c$) upon a successful entangling gate remove the unwanted intermediate qubits with a $Y$or $Z$  measurement, depending on the entangling gate; ($d$) a 2D honey-comb cluster state is produced, which is universal for quantum computing. }
 \label{fig:2dcluster}
\end{figure}

Let us denote the Pauli matrices $\sigma_x$, $\sigma_y$, and $\sigma_z$ by $X$, $Y$, and $Z$, respectively. It is a general property of cluster states that an $X$-measurement of a qubit in a chain both shortens the chain and creates a cherry, as shown in Fig.~\ref{fig:2dcluster}(a,b). We can now apply the entangling gate to the cherries of two logical qubit chains. When the gate fails, we can perform a $Z$-measurement on both cherries to purify the cluster state.  Both chains will be shortened by two qubits. Note that this procedure places the restriction $c_2 \leq 1$ on the entangling gate, which rules out the fusion gates. When the entangling gate is successful, the resulting cluster fragment is shown in Fig.~\ref{fig:2dcluster}(c). Depending on the particular action of the entangling gate, $Y$- or $Z$-measurements of the two qubits between the chains will remove them from the cluster and maintain a single vertical edge. The procedure can be repeated to create the two-dimensional honeycomb lattice cluster of Fig.~\ref{fig:2dcluster}(d). This two-dimensional cluster state is universal for quantum computing \cite{vdnest06}.

There are now two questions: how costly is this procedure in terms of the necessary number of qubits in the horizontal cluster chains $\braket{N}$, and how does this procedure affect the random walk fluctuations $\Delta N$?

\begin{figure}[t]
 \includegraphics[width=8.5cm]{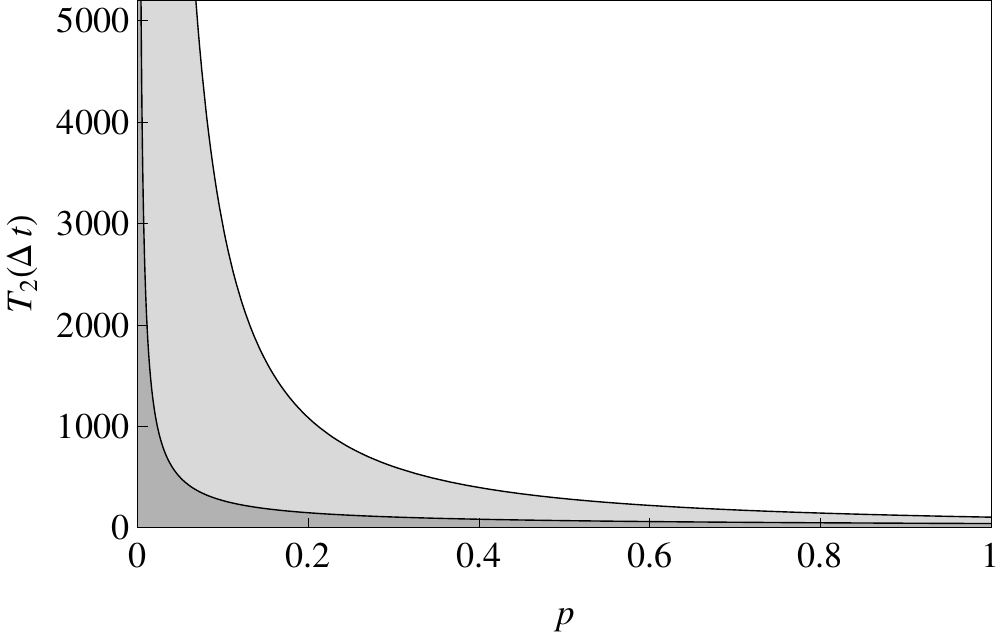}
 \caption{Threshold of the $T_2$ lifetime of the qubits in a two-dimensional cluster state as a function of the entangling gate success probability $p$ with $\alpha=10$ for double heralding (upper curve) and the broker-client model (lower curve).}
 \label{fig:t22d}
\end{figure}

To answer the first question, we note that each attempt shortens the horizontal chain by two qubits. On average, we have to attempt the entangling gate $1/p$ times (the cluster may already have cherries present, so this is a worst-case scenario). Since each qubit in the cluster will have exactly one vertical edge, the total buffer length must increase by this factor:
\begin{align}
 \braket{N}_{\rm 2D} = \frac{2}{p}\beta\Delta N = \frac{2}{p}\, \sqrt{2\ln\alpha}\;\Delta N\, .
\end{align}
The fluctuations $\Delta N$ due to the adding of the mini-clusters of size $m$ do not change. When we take this into account, the $T_2$ threshold for two-dimensional honeycomb cluster states becomes
\begin{align}\label{eq:threshold2d}
 T_2 & = \alpha \left( \tau + \frac{2\beta}{p}\sqrt{\frac{(1+c_2)^2}{p} - c_2(2-c_2)} + \frac{1+p c_1 +(1-p)c_2}{p} \right) \cr
 & = \alpha\left( \tau + 4\sqrt{\frac{(4-p)\ln\alpha}{p^3}} + \frac{2}{p} \right) \quad\text{for}~ c_1=c_2=1\, .
\end{align}
The behaviour of the $T_2$ time for the double-heralding and the broker-client model is charted in Fig.~\ref{fig:t22d}. The increased number of  entangling gates predictably give rise to a longer $T_2$ threshold, particularly when $p$ is small. However, it should be noted that these values are still within practical limits for reasonably small (but not tiny) $p$. Note also that the broker-client model performs significantly better for two-dimensional cluster states. This is due to the fact that the ``client'' qubits in the brokering scheme are not at risk, and the vertical edges can be attempted without shortening the buffer. In general, the capabilities for creating cluster states increase dramatically when we have access to more than one qubit per site, provided these qubits allow deterministic two-qubit gates between them. 

When $c_2>1$, the above method for creating two-dimensional clusters does not work, and an altogether more complicated (and more costly) cluster state must be prepared, which allows us to create the vertical links while protecting the horizontal edges. This means that the fusion gates are not ideal for this strategy of creating higher-dimensional cluster states.

\begin{figure}[t]
 \includegraphics[width=8.75cm]{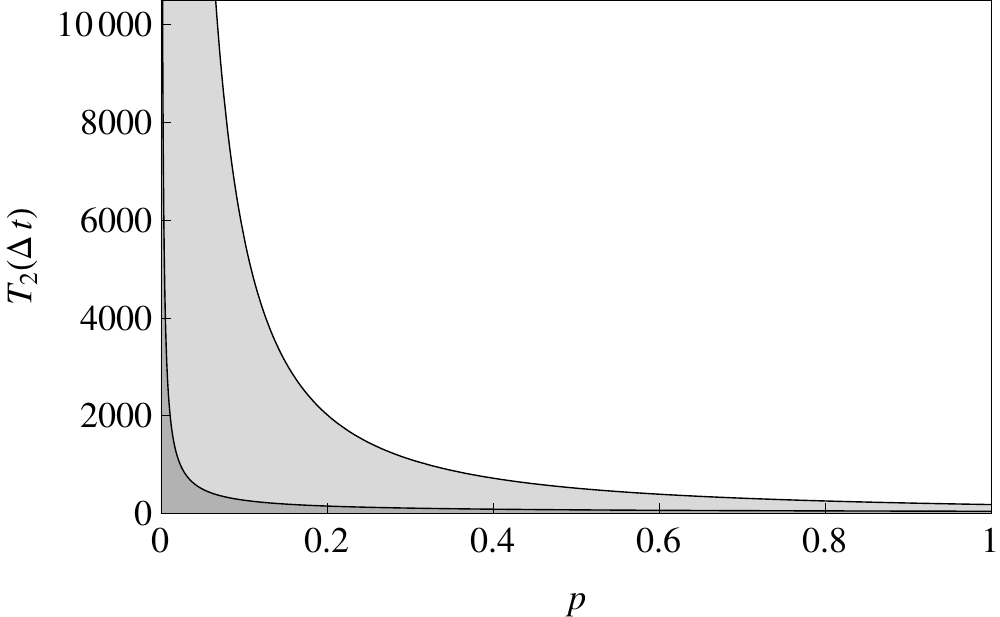}
 \caption{Threshold of the $T_2$ lifetime of the qubits in a three-dimensional cluster state as a function of the entangling gate success probability $p$ with $\alpha=10$ for double heralding (upper curve) and the broker-client model (lower curve).}
 \label{fig:t23d}
\end{figure}

A similar approach can be used to create three-di\-men\-sion\-al clusters. The scaling behaviour is very close to that of the two-dimensional cluster, since we still operate on the assumption that each physical qubit has only one extra edge in addition to its two horizontal edges. Therefore, instead of a factor $2\beta/p$ we have a factor $4\beta/p$: 
\begin{align}\nonumber
 \frac{T_2}{\alpha} = \tau + \frac{4\beta}{p}\sqrt{\frac{(1+c_2)^2}{p} - c_2(2-c_2)} + \frac{1+p c_1 +(1-p)c_2}{p} \, .
\end{align}
It is easy to see that this argument generalises to arbitrary dimensions $d\geq 2$. Again, the broker-client model does not pay this extra penalty. The scaling behaviour of $T_2$ as a function of $p$ is given in Fig.~\ref{fig:t23d}.

\begin{figure}[b]
 \includegraphics[width=8.6cm]{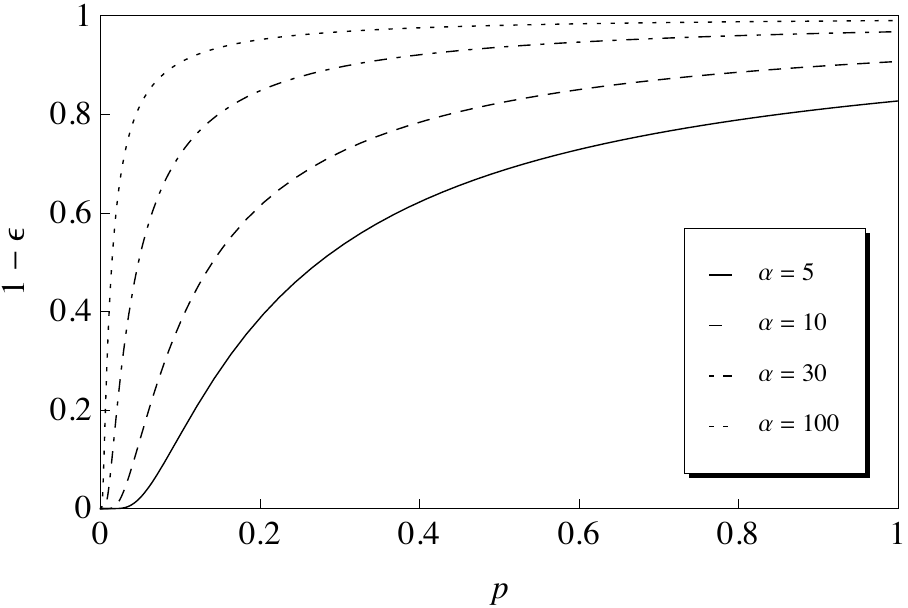}
 \caption{For given fault tolerance parameters $\alpha$, the probability of \emph{not} accumulating a bit flip error $1-\epsilon$ in higher-dimensional cluster states drops sharply with decreasing $p$.}
 \label{fig:bitflip}
\end{figure}

One complication with this method of creating higher-dimensional cluster states using entangling operations that do not have a broker-client structure is that the main cluster can accumulate bit flip errors that are teleported in by failed entangling operations between different linear cluster chains (the vertical edges in Fig.~\ref{fig:mbqc}). The probability $\epsilon$ of a bit flip error increases with each measurement of a failed entangling operation, and must itself be below the fault tolerance threshold. We need to make on average $2/p$ qubit measurements per vertical edge, and each measurement has a small probability $\delta$ of teleporting a bit flip error into the linear cluster chain. This probability is related to the time $t$ in the cluster via
\begin{align}
 \delta = \frac12 \left(1-e^{-t/T_2}\right) .
\end{align}
The choice $t = \tau_{\rm av} = T_2/\alpha$ places an upper bound on $\delta$ that is independent of $T_2$. The total probability of \emph{not} having a bit flip error is then calculated as 
\begin{align}\label{eq:bitflip}
 1-\epsilon = (1-\delta)^{2/p} = e^{-\frac{1}{\alpha p}} \; \cosh^{\frac{2}{p}}\left( \frac{1}{2\alpha} \right) .
\end{align}
The value of $1-\epsilon$ as a function of the success probability of the entangling operation is shown in Fig.~\ref{fig:bitflip}. Since $\cosh x \geq 1$ for all $x$, Eq.~(\ref{eq:bitflip}) leads to the requirement that $\alpha \gg 1/p$. Note that this argument applies only to entangling operations that act directly on the qubits in the cluster, and not to entangling operations that operate according to the Broker-Client model.

Finally, it is instructive to compare the ratios found here with practical quantum systems, such as NV centres in diamond. In the case of the broker-client model \cite{benjamin06}, the ratio between the $T_2$ time and the gate time $\Delta t$ is of the order of $3\times 10^3$ when the success probability is about $1\%$. This model requires two-qubit entangling gates between the nuclear and electron spin of an NV centre, which has a duration of about 0.1~\micro s \cite{jelezko04}. The entangling operation between electron spins in qubits may be on the order of 1~\micro s, and the $T_2$ time for NV centres can be up to 10~ms \cite{wrachtrup06}. This indicates that the just-in-time Broker-Client method for quantum computing with cluster states is a promising candidate for the practical construction of a quantum computer.

\section{Size of the Qubit Reservoir}
\noindent
Apart from the time it takes the qubits from first initialisation to measurement, a very practical consideration is the size of the qubit reservoir necessary to produce the mini-clusters of size $m$. Previous proposals minimised the time that a qubit must remain coherent at the expense of a qubit reservoir that grows exponentially in $p$. However, given the difficulty of creating a large number of identical qubits, this will most likely not be feasible. 

\begin{figure}[t]
 \includegraphics[width=8.5cm]{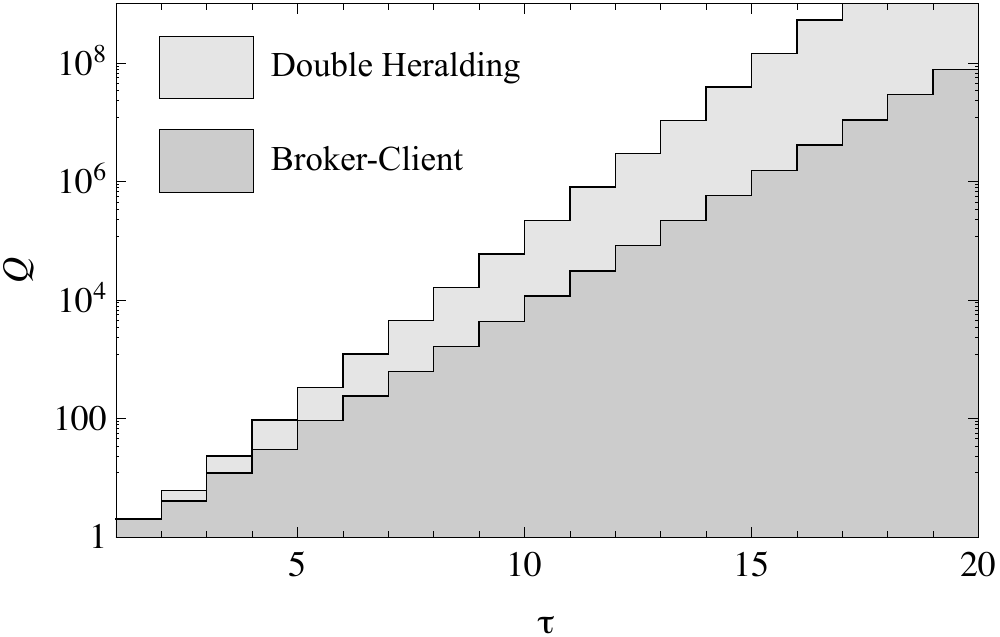}
 \vskip-3mm
 \caption{The required size of the reservoir $N$ that produces on average one mini-cluster of length $m$ as a function of the probability $p$. The number of time steps $\tau$ allowed for creating the mini-clusters is $\tau = 1/p$. The reservoir increases exponentially with $1/p$.}
 \label{fig:reservoir}
\end{figure}

We require that the size $Q$ of the reservoir per logical qubit is such that after $1/p$ steps there is on average one linear mini-cluster of length $m$ given in Eq.~(\ref{eq:minicluster}). The total size of the reservoir is therefore $M\times  Q$, and each logical qubit has an attempted cluster increase at each step in the computation. The larger $M$, the more likely it is that there are close to $M$ mini-clusters in total. We made a numerical estimate of the reservoir size $Q$ for the Double Heralding and the Broker-Client protocol, shown in Fig.~\ref{fig:reservoir}. There are several strategies that can be employed in creating the mini-clusters, such as {\sc greed}, {\sc modesty}, or {\sc random} \cite{gross06}. These strategies provide rules how small clusters are combined in the pursuit of the mini-clusters of length $m$. We used a {\sc random} strategy, where entangling operations are attempted between randomly chosen pairs of cluster states. This minimised the computational time. We expect that the performance of the {\sc random} strategy lies somewhere between {\sc greed} and the near-optimal strategy {\sc modesty}. We find that the size of the reservoir still grows exponentially with the number of times steps $\tau = 1/p$. A least-square fit on the numerical data allows us to extrapolate the scaling, and we find that  
\begin{align}
 Q_j \simeq \exp(\gamma_j\tau)\, , 
\end{align}
where $j$ denotes Double Heralding (DH) or Broker-Client (BC), and the fitting parameters $\gamma_j$ are given by 
\begin{align}
\gamma_{\rm DH} = 1.30\qquad\text{and}\qquad  \gamma_{\rm BC} = 0.979\, . 
\end{align}
This assumes that each qubit can engage in an entangling operation with any other qubit in the reservoir. Fig.~\ref{fig:reservoir} shows that the size of the reservoir quickly becomes prohibitively large, and the success probability of the entangling operation should not be much lower than 20\%.

\section{Conclusions}
\noindent
We have derived expressions for the minimum qubit coherence time necessary for maintaining a sustained quantum computation using the just-in-time measurement based architecture for one-, two-, and three-dimensional cluster states. It is assumed that the qubits are cheap, such that it is easy to create a large number of identical qubits. In addition, the overhead that is needed to attempt an entangling gate between two arbitrary qubits in the reservoir must also be manageable. We find that the ratio of the qubit $T_2$ coherence time to the gate time is within practical limits. This estimate of the required coherence time has assumed that single qubit gates and measurements are fault-free, a condition that will most certainly not be satisfied in any realistic application. The simple arguments here give a lower bound on the $T_2$ time of the qubit. The size of the qubit reservoir that created the mini-clusters scales exponentially with $1/p$, which places a practical limit on the success probability of the entangling operation.

\section*{Acknowledgments}
\noindent
PK wishes to thank Simon Benjamin, Dan Browne, Earl Campbell and Joe Fitzsimons for fruitful discussions. This paper is dedicated to the memory of Sean D. Barrett.

\end{document}